\documentstyle[12pt,epsf]{article}
\newcommand{\beq}{\begin{equation}}
\newcommand{\eeq}{\end{equation}}
\newcommand{\ppd}{\partial}

\title{Bag formation in Chiral Born-Infeld Theory }

\author{ O.V. Pavlovsky \thanks{e-mail address:
ovp@goa.bog.msu.su }  \\ {\em The Abdus Salam  International
Centre for Theoretical Physics, Trieste, Italy} \\
{\em and} \\
{\em
Bogoliubov Institute for Theoretical Problems of Microphysics,} \\
{\em Lomonosov Moscow State University } \\ {\em Moscow, 119992,
Russian Federation.} }

\date{ \ \ \  }

\begin{document}

\maketitle

\begin{abstract}
The "bag"-like spherically symmetrical solutions of the Chiral
Born-Infeld theory are studied. The properties of these solutions
are obtained, and a possible physical interpretation is also
discussed.
\end{abstract}

\vspace{1cm}

PACS number(s): 12.39.Dc, 12.39.Fe, 05.45.Yv, 11.27.+d, 11.30.Qc

\section{Introduction}
The construction of the low-energy baryon state model is a
long-standing task, and perhaps the  first realistic solution of
this problem was proposed by Skyrme \cite{skyrme}. Within the
Skyrme approach baryon treated as a topological soliton of
non-linear chiral meson field and the topological number of such
soliton is associated with the baryon number. This idea was very
popular in the eighties due to the intense interest to the
solitonic physics at that time. The Skyrme model is well studied
and gives us the quite good description of the many low-energy
nuclear physics phenomena.

On the other hand,  from the DIS experiments  we know, that the
baryon consists of the charged constituents (so called constituent
quarks), but the experiments with  spin of proton showed that only
twenty percents of the baryon spin can be associated with the
charged constituents \cite{spin}! Therefore, the  realistic model
for low-energy baryon state should describe the quark constituents
as well as the chiral degrees of freedom of the meson cloud around
them.

But how to combine these two opposite points of view on the baryon
into the  framework of a one model? In order to find the model
with such dualistic properties, the so-called "two-phase" models
were proposed. The simplest realization of the two-phase idea is
the well-known Chiral Bag Model (ChBM) \cite{bag}.  In this model,
the constituent quarks are placed inside a spherical region
(so-called "bag"), subject to the confining boundary conditions,
and meson cloud lives outside. As a rule, the exterior pion field
are Skyrmionic but the quark part is necessary in this model  for
the self-consistency of the model, as will as for the
phenomenological point of view. Typical Lagrangian of the
two-phase model consists of the three parts
\begin{equation}\label{1.0}
{\cal L}_{\mbox{model}} =  {\cal L}_{\mbox{q}}\theta_V  + {\cal
L}_{\pi} \theta_{\bar V} + {\cal L}_{\partial V} \delta_V
\end{equation}
where Heaviside's $\theta_V$ function cuts out the internal region
for quark part of the Lagrangian $L_q$, inverse $\theta_{\bar V}$
function cuts out the external region for meson field with the
Lagrangian $L_\pi$.  The crucial point of this model steams  from
the fact that the kinematic consequence of the confinement
boundary conditions for fermions in the bag is the non-zero axial
current across the boundary. For self-consistency of the model,
these two regions should be joined by the requirement of
continuity of the axial current on the boundary. In order to
guarantee such continuity, the boundary term $L_{\ppd V}$ must be
introduced.

Chiral Bag Model is used very widely in the phenomenology of
elementary particle physics \cite{phen} but on the other hand this
model has serious methodological  disadvantages. First of all, as
one can easily see in (\ref{1.0}), this Lagrangian depends
directly  on the geometry of the confinement region $V$. In the
case of the Chiral Bag Model, $V$ is a sphere. But it would be
very interesting to find a theory where the form of the
confinement region is generated dynamically, rather than is chosen
"by hand", as in the case of the ChBM. Moreover, the geometry
dependence of the Lagrangian (\ref{1.0}) directly violates the
Lorenz covariance of the theory, and thus there is no a way for
connection such complicated and non-invariant construction with
QCD.

 \begin{figure}[t]
 \leavevmode
 \epsfbox{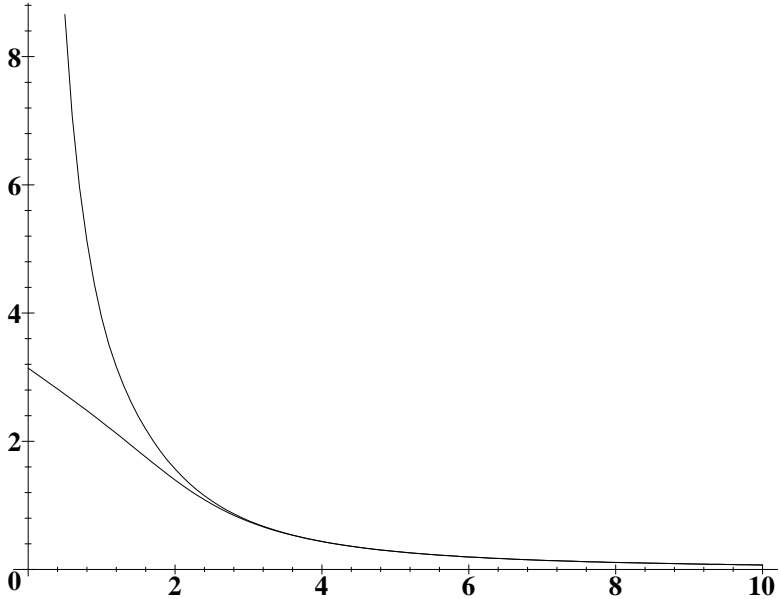}
 \caption{Singular solution of prototype theory (2) and soliton solution of Skyrme model (skyrmeon).
  Horizontal axis: r (in fm).}
 \end{figure}

Another kind of methodological difficulties appears from Skyrme
part of the Skyrme Model \cite{skyrme}. The point is that the simplest
realization of the non-linear meson field (so-called Weinberg's
prototype theory) \begin{equation}\label{1.1} {\cal
L}_{\mbox{pr}}= \frac {f_\pi^2}{4} {\rm Tr} L_\mu L^\mu
\end{equation}
has no soliton solution and any non-trivial static solutions of
this theory have infinite self-energy. But on the other hand,
the theory (\ref{1.1}) correctly describes the behavior of the meson
field at large distance from the source (the proton or neutron).
In order to regularize the singular self-energy of the prototype
theory, Skyrme proposed to modify slightly the theory
\cite{skyrme} by means of the additional term ${\cal
L}_{\mbox{Sk}}={\rm Tr} [ L_\mu
 , L_\nu ]^2/e$ (see Fig. 1). Such term in the Lagrangian of the meson field leads
to generation of the soliton solutions (skyrmions), and now is
treated as the leading order of the expansion of the effective
meson theory in strong limit (by analogy with the Chiral
Perturbation Theory Lagrangian defined as series). But
unfortunately such analogy is not so good for the soliton paradigm
of baryons. Basing on the renormalization group approach, one can
show that the coefficients of such expansion depended strongly on
the energy scale of the process, and in the strong regime (near
the nucleon) the highest terms of the expansion become more and
more essential \cite{renorm}. This fact is connected with the
nonrenormalizable nature of chiral field theory and this situation
is similar with quantum gravity.

An additional point to emphasize is that the Skyrme regularization
procedure does not solve the problem of singular solutions of
chiral field theory, because Skyrme model has a lot of another
singular solutions along with soliton sector.

 \begin{figure}[t]
 \leavevmode
 \epsfbox{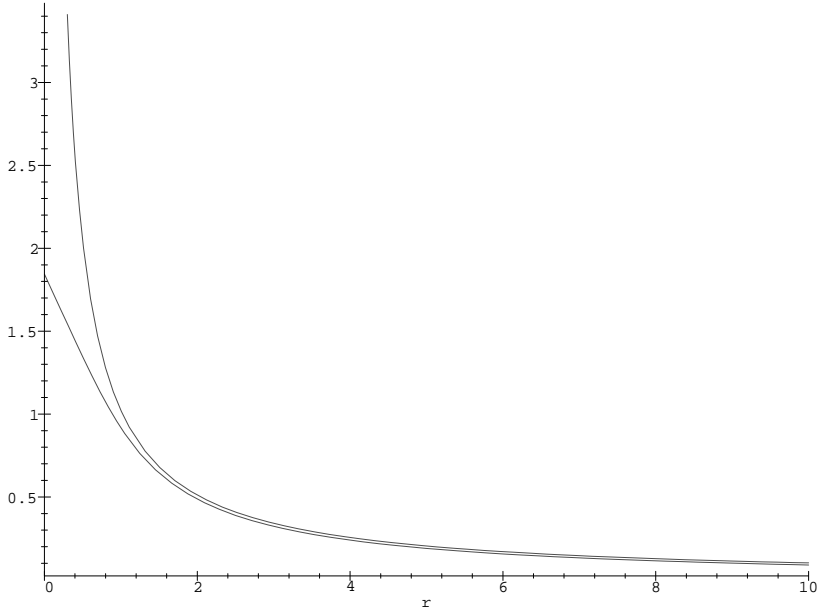}
 \caption{Singular electrostatic Coulomb potential $-\phi(r)$ and soliton solution of Born-Infeld theory (3).
  Horizontal axis: r (in fm).}
 \end{figure}

In order to construct a suitable candidate for the role of the
chiral effective theory, let us consider a very illuminative
historical analogy with the problem of singular self-energy of
electron. Before the Era of Quantum Electrodynamics, Born and
Infeld \cite{born} proposed the non-linear covariant action for
the electro-magnetic fields with very attractive features.
Firstly, in the framework of the BI theory, the problem of
singular self-energy of electron can be solved. In this theory the
electron is a stable finite energy solution of the BI field
equation with electric charge. Second, the BI action has the scale
parameter $\beta$. Using expansion in this parameter, one reduces
the BI action to the usual Maxwell form in the low-energy limit.
Indeed, using the analogy with the relativistic particle action,
let us consider the action in the Born-Infeld form
\begin{equation}\label{1.2}
   {\cal L}_{BI} = -
 \beta^2 \bigg(1-\sqrt{1+\frac 1{\beta^2}F_{\mu \nu} F^{\mu \nu} } \bigg)
\stackrel{\beta \to \infty }{\longrightarrow} \frac{1}{2} F_{\mu
\nu} F^{\mu \nu}
\end{equation}

 It can be shown that the solution for the electrostatic spherically
 symmetrical potential $ A_0=\phi $ reads

\begin{equation}\label{1.3}
    \phi(r,\beta) = C \int^R_0{\frac{dr}{\sqrt{r^4+C^2/\beta^2}}}
 \stackrel{R \to \infty }{\longrightarrow} - \frac{C}{R} + \underline{O} ( \frac{1}{R^5} )
\end{equation}

In Fig. 2 the Coulomb solution and the relevant soliton of BI
electrodynamics are presented. Comparing Fig.1 and Fig. 2, one
finds that the situations with the singular salfenergy in these
two theories (singular selfenergy of the Coulomb potential and
singular selfenergy of the solutions of the prototype Lagrangian
(\ref{1.1}) ) are very similar and in our paper we apply the very
similar procedure for the chiral prototype Lagrangian (\ref{1.1}).

\section{Chiral Bag solution of Chiral Born-Infeld Theory}

Let us consider  the Lagrangian
\begin{equation}
{\cal L}_{ChBI} = - f^2_\pi {\rm Tr} \beta^2 \bigg(1-\sqrt{1-\frac
1{2\beta^2}L_\mu L^\mu } \bigg) \stackrel{\beta \to \infty
}{\longrightarrow} - \frac{f^2_\pi}{4} {\rm Tr} L_\mu L^\mu ,
\label{2.1}
\end{equation}
where $\beta$ is the mass dimensional scale parameter of our
model. It can be easily shown that the expansion of the Lagrangian
(\ref{2.1}) gives us the prototype theory as the leading order
theory in the parameter $\beta$. The theory (\ref{2.1}) is the
direct analogue of the Born-Infeld action for chiral fields.

Our theory contains only the second-order derivative terms. Thus
the dynamics of this theory can be studied in detail. Now we
consider the spherically symmetrical field configuration
\begin{equation}
U=e^{i F(r) (\vec n \vec \tau)}, \qquad \vec n = \vec r/|r| .
\label{2.2}
\end{equation}
The energy of such configuration is the functional
\begin{equation}
E^\beta [F] = 8 \pi f^2_\pi \beta^2
\int\limits_0^\infty \Bigl( 1 - R \Bigr) r^2 dr ,
\label{2.5}
\end{equation}
where
$$
R=\sqrt{ 1-\frac 1{\beta^2}\Bigl(\frac{F'^2}{2} +  \frac {\sin^2
F}{r^2}  \Bigr)
 } .
$$

Using the variation principle, we get the equation of motion
\begin{equation}
\Bigl(r^2 \frac{F'}{R} \Bigr)' =  \frac{\sin 2F}{R}
\label{2.3}
\end{equation}
and for amplitude $F(r)$ we obtain
$$
(r^2 - \frac 1{\beta^2} \sin^2 F) F'' + ( 2rF' -\sin 2F) -
$$
\begin{equation}
- \frac 1{\beta^2} (r F'^3 - F'^2 \sin 2F + 3 \frac 1{r} F' \sin^2F
- \frac 1{r^2} \sin2F \sin^2 F)=0.
\label{2.4}
\end{equation}

The next aim of our investigation is finding the solutions of
equation (\ref{2.4}). This equation  is a very complicated
nonlinear differential equation. In order to solve it, only
numerical or approximation methods seem applicable. The crucial
point of such analysis is that the leading derivative term in this
equation contains the factor
\begin{equation}
\Bigl( r^2 - \frac{1}{\beta^2} \sin^2 F \Bigr).
\label{2.6}
\end{equation}

Due to this factor, equation (\ref{2.4}) has a singular region
(singular surface) with singular behavior of solutions. One can
find an equation on such surface by using the standard singular
perturbation theory techniques \cite{singular}. Let $ r_0 $ belong
to such singular surface. Then
\begin{equation}
\left \{  \begin{array}{rcl}
            (\beta r_0)^2  - \sin^2 F_0 &=& 0\\
F'_0 \Bigl( (F'_0)^2 \sin F_0 \mp F'_0 \sin 2F_0 + \sin^2 F_0
 \Bigr) &=& 0 \, \, \, \, \mbox{where} \, F_0=F(r_0) .
           \end{array}
\right.
\label{2.61}
\end{equation}
Equations (\ref{2.61}) have only two solutions: whether
\begin{equation}
\left \{  \begin{array}{lcl}
           r_0 &\neq& 0 , \\
           F_0 &=& \pm \arcsin(\beta r_0) + \pi N   , \, \, \, \mbox{where} \,
 N \in \mbox{\textbf{Z}}, \\
          F'_0 &=& 0.
          \end{array}
\right. \label{2.62}
\end{equation}
or
\begin{equation}
\left \{  \begin{array}{lcl}
           r_0 &=& 0 , \\
           F_0 &=&  \pi N   , \, \, \, \mbox{where} \, N \in
           \mbox{\textbf{Z}}, \\
           F'_0 &\neq& 0.
           \end{array}
\right. \label{2.63}
\end{equation}

Topological solitons of ChBI theory  correspond to the possibility
(\ref{2.63}) were studied in \cite{soliton}. In our paper we
consider the more interesting possibility (\ref{2.62})

Using the standard singular perturbation theory procedure
\cite{singular}, one obtains the asymptotic behavior near the
singular surface ($r \to r_0$, $F(r \to r_0) \to \arcsin(\beta
r_0)$)
\begin{equation}
F(r \to r_0) = \arcsin(\beta r_0) + \mbox{\textbf{b}} (r -
r_0)^{3/2} + \underline{O}((r - r_0)^{2}), \label{2.64}
\end{equation}
where $\mbox{\textbf{b}}$ is a constant. Of course, the derivative
$F'_0$ at the point $r=r_0$ is zero.

To guarantee the finiteness of the energy of our solutions, we
should choose the following asymptotic at infinity ($r \to
\infty$)
\begin{equation}
F(r) = a_\infty \,
(1/r)^2 - \frac{a_\infty^3}{21} \, (1/r)^6 -
\frac{a_\infty^3}{3\beta^2} \, (1/r)^8 + \underline{O}(1/r^{10}),
\label{2.8}
\end{equation}
where $a_\infty$ is a constant.

Notice that equation (\ref{2.4}) has very useful symmetries.
First of all, this equation is symmetrical with respect to the
changes $F \leftrightarrow F+N\pi, \, N \in {\bf Z} $ and
$F \leftrightarrow -F$.

 \begin{figure}[t]
 \leavevmode
 \epsfbox{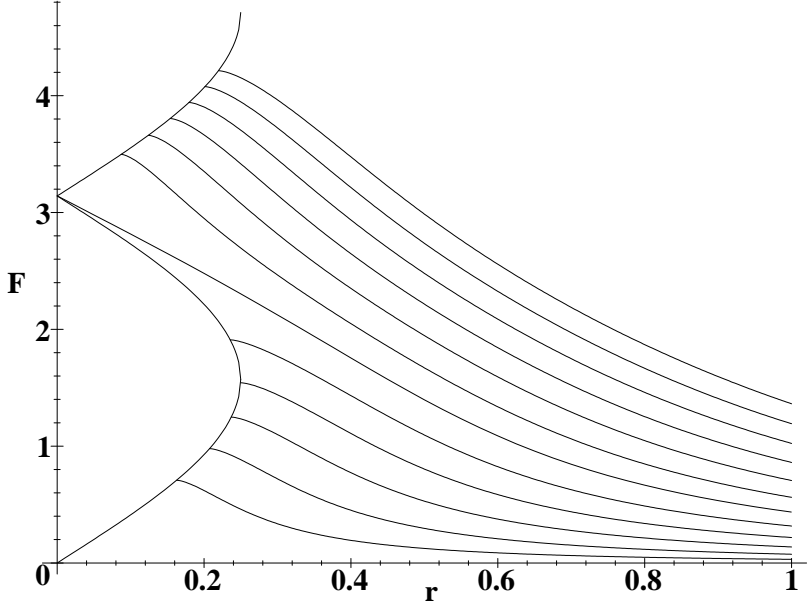}
 \caption{ Solutions of equation (9)
which have the asymptotics (15) ($a_\infty>0$) at infinity.
Horizontal axis: r (in fm).}
 \end{figure}

 \begin{figure}[t]
 \leavevmode
 \epsfbox{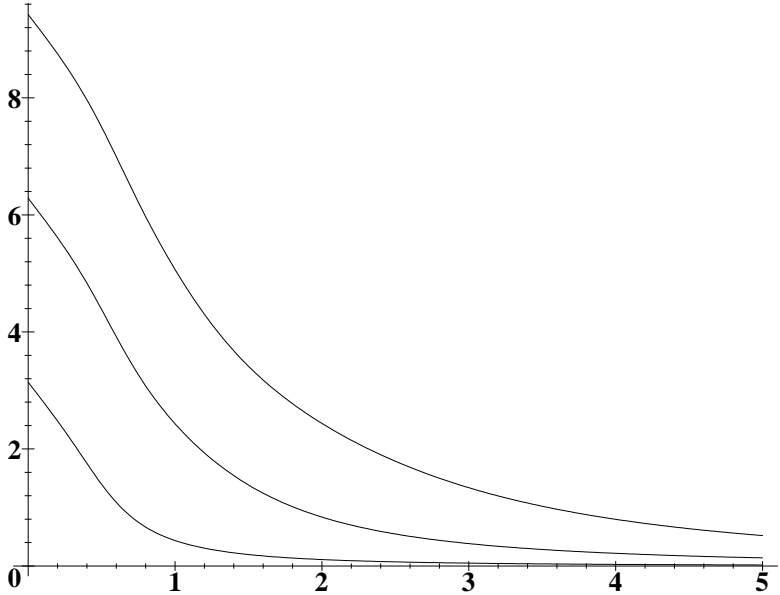}
 \caption{Solitons with $B=1, 2$ and $3$.
  Horizontal axis: r (in fm).}
 \end{figure}


The numerical investigation of the solutions of equation
(\ref{2.4}) which have the asymptotics (\ref{2.62}) ($a_\infty>0$)
at infinity is presented in Fig.3. Most of these solutions can be
evaluated only for $r>r_0$, where $r_0$ is determined by $F(r_0)=
\pm \arcsin (\beta r_0)$. But among this set of solutions there
are solutions with $r_0=0$. Such solutions have the asymptotics
\cite{soliton}
\begin{equation}\label{2.66}
F(r) = \pi N  + a r -
\frac{7a^2-4\beta^2}{30(a^2-\beta^2)} a^3 r^3+
\underline{O}(r^{5})
\end{equation}
at origin ($a^2 < \beta^2/3$ is a constant), and these are the
topological solitons of the ChBI theory. Solitons with $B=1, 2$
and $3$ are presented in Fig.4. The scale parameter $\beta=807 \,
\mbox{MeV}$ is preliminarily defined from the hypothesis that the
soliton with $B=1$ is a nucleon.

Now we would like to draw the attention to another class of
solutions. These solutions are defined everywhere, except the
small ($ \sim$ 0.2 fm) spherical region about the origin. These
solutions look like a ''bubble" of vacuum in the chiral fields and
are of the interest for the chiral bag model of baryons
\cite{bag}. In the internal region ($r<r_0$), the only vacuum
configuration can exist. From the mathematical point of view, such
"step-like'' solutions with jump of $F(r)$ at $r=r_0$ are a
generalized solutions \cite{general}.

To clarify the physical nature of this "step-like'' generalized
solutions, let us consider the projection of the left-hand chiral
current on the outward normal of the singular surface. As we
pointed out in the Introduction, a non-zero chiral current across
the boundary the kinematic is the consequence of the confinement
boundary conditions for the fermions in the bag. According to our
proposition, the singular surface of "step-like'' solutions is a
confinement boundary surface for constituent quarks. For the
self-consistency of our "two-phase" picture, let us check is it
possible to compensate the non-zero quark chiral current on the
confinement surface by the non-zero chiral current of ChBI
"step-like'' configuration that appear due to the defect on the
singular surface. The projection the chiral current on the outward
normal of the singular surface for the spherically symmetrical
configuration (\ref{2.2}) reads
\begin{equation} \label{2.71}
(\vec n \, \vec J_\pi )|_{\partial V}= f^2_\pi {\rm Tr}
\frac{\tau^a  \vec x  \vec L }{\sqrt{1-\frac 1{2\beta^2}L_\mu
L^\mu }} \, = \, \frac{3}{2} f^2_\pi \frac{\mbox{\textbf{b}}
r_0}{\sqrt{2 r_0 + 9 r_0^2 \mbox{\textbf{b}}^2 }},
\end{equation}
where the coefficient $\mbox{\textbf{b}}$ from the asymptotic
(\ref{2.64}) is a function of $F_0=F(r_0)$ and can be evaluated
numerically.

The crucial point for such analysis  steams  from the fact that
$\mbox{\textbf{b}}(F_0)$ has the singularity at $F_0=\pi N$ and
$r_0=0$, or $\mbox{\textbf{b}}(F(r_0=0))=\infty$. This implies
that the soliton solutions of the ChBI theory are solutions with
point-like singular chiral source and the "bag"-like solutions are
the solutions with the some distribution of the chiral current on
the confinement surface. It is possible to show that for any
internal constituent quark configuration with confinement inside
some volume $V$ the solution of ChBI theory $U( \vec \pi)$ could
be defined which compensate the non-zero quark's chiral current
across the surface $\partial V$
\begin{equation}\label{2.72} (\vec n \, \vec J_q )|_{\partial V} =
\sum_g{\frac{i}{2}\bar{\Psi}_q (\vec{x}\vec{\gamma}) \gamma_5
\Psi_q} = (\vec n \, \vec J_\pi )|_{\partial V}
\end{equation}
Equation (\ref{2.72}) can be considered as a condition on
coefficient $\mbox{\textbf{b}}(\partial V)$, and plays the role of
the self-consistency condition between quark and chiral phases.

In conclusion of this section, it would be interesting to point
out that the condition (\ref{2.72}) is not jast an artifact of the
spherical symmetry of our task. This is a direct consequence of
the asymptotics (\ref{2.64}) but, as easy to see, such asymptotics
can depended only on local characteristics of surface $\partial
V$. It means that the asimptotics of the field configuration along
the normal in the leading order depends only on the curvature
radius of the confinement surface at this point and equation
(\ref{2.72}) can be used for any other geometry of $\partial V$.

\section{Conclusions}
The aim of this paper is to study of  "bag"-like generalized
solutions of the Born-Infeld theory for chiral fields. These
solutions  were firstly proposed in our work \cite{soliton}. For
simplicity, in this work we restricted ourselves by the
spherically symmetrical configuration (\ref{2.2}), but, of course,
the ChBI theory has solutions with another geometry of the
singular surface. Physically the critical behavior in the ChBI
theory appears when the chiral field strength of the prototype
field theory approaches the value of the squarred f effective
coupling constant ($\beta^2$). This is a very physical idea. Today
we know that there are many theories which have the low-energy
limit in the form of the Born-Infeld action \cite{BIstrings}.

The Chiral Born-Infeld theory is a good candidate on the role of
the effective chiral theory and the model for the chiral cloud of
the baryons. In this model one can find not only spherical "bags",
it is possible to also study the "string"-like, toroidal or
"\textbf{Y}-Sign"-like solutions, or some another geometry. The
geometry of confinement surface  depends directly on the model of
color confinement and it would be very interesting to use, for
example, the Lattice QCD simulations for the color degrees of
freedom in combination with our model for external chiral field.

In conclusions I would like to point out that there are many
interconnections between Lattice QCD at strong coupling limit and
ChBI theory. It is possible to prove  such low-energy chiral limit
of QCD using the lattice regularization of QCD. All this questions
should be the topics for future works.

\vspace{1.5cm}

I would like to thank Dr. Igor Cherednikov for interesting
discussions and comments. This work is partially supported by the
Russian Federation President's Grant 1450-2003-2. The hospitality
and financial support of the ICTP in Trieste is gratefully
acknowledged.

\end{document}